\documentstyle[aps]{revtex}
\begin{document}
\title{Consistency of Wilsonian effective actions}
\author{Vipul Periwal}
\address{Department of Physics,
Princeton University,
Princeton, New Jersey 08544}

\def\dd{\hbox{d}}
\def\tr{\hbox{tr}}\def\Tr{\hbox{Tr}}
\maketitle
\begin{abstract} Wilsonian effective actions are interpreted as 
free energies in
ensembles with prescribed field expectation values and prescribed   
connected two-point functions.
Since such free energies are directly obtained from
two-particle-irreducible functionals, it follows that Wilsonian 
effective actions satisfy elementary perturbative consistency 
conditions,
and non-perturbative convexity conditions.
In particular, the exact determination of a Wilsonian action by other
means (e.g. supersymmetry) 
allows one to extract restrictions on the particular cutoff scheme and field
reparametrization that would lead to such a Wilsonian action from an
underlying microscopic  action.  
  
\end{abstract} 

\def\al{\alpha}
\def\be{\beta}
\def\eps{\epsilon}
The Wilsonian effective action\cite{wilson}
figures prominently in the spectacular 
advances\cite{sw,seiberg} 
that have been made in the understanding of the dynamical 
properties of supersymmetric quantum field theories.  This kind of 
effective action does not involve integrating out massless degrees 
of freedom, hence has locality properties that allow one to utilize
properties of supersymmetric actions such as holomorphy that do not
hold for the 1PI effective action when massless particles are present.
Of course, the Wilsonian action is to be used in a functional 
integral over the remaining long wavelength modes, unlike the 1PI action.
Based on such properties, remarkable
arguments have been given to determine the low-energy dynamics of
supersymmetric field theories\cite{reviews}. There appears to be great 
predictive power in this approach.

Given the physical importance of   the Wilsonian 
effective action, it is natural to ask if there is a 
regulator-independent physical characterization of 
the Wilsonian action, analogous to
the interpretation of the 1PI effective action as the free energy in
an ensemble with a prescribed field expectation value.  This question 
becomes especially interesting in light of the 
unease that has been expressed\cite{dine}   regarding 
the holomorphy assumptions that play a large part in the 
supersymmetric applications mentioned above.  
Ref.~\cite{dine} argued, following Shifman and Vainshtein\cite{shif}, 
that not just
the microscopic action, but also the exact `field-dependent cutoff'
scheme and possible field reparametrizations, must be regarded as 
additional parameters in low-energy Wilsonian actions.  This would 
seem to open up a veritable Pandora's box of arbitrariness in 
deductions based on Wilsonian effective actions.   

It would appear to follow  then that the
exact determination of a low-energy Wilsonian action should allow
the determination of the cutoff scheme, and  the field
reparametrization, that leads from the microscopic action to the
Wilsonian action.  In particular, if one has deduced the exact
non-derivative terms in the Wilsonian action, one might be able to 
construct (or at least find restrictions on the
form of) the rest of the Wilsonian action from such knowledge.  This 
is, potentially, of great interest for a variety of reasons (see below).
Unfortunately, one cannot carry this program out for lack of a 
scheme-independent characterization of the Wilsonian action.

The aim of this note is to present a simple and explicit scheme for 
extracting such information, by realizing that an alternative 
physical characterization of the Wilsonian action can be given in terms of
the free energy in a particular ensemble.  This physical 
interpretation is non-perturbative, and makes no explicit reference to
any cutoff scheme.  It therefore places the Wilsonian action in a
context that allows one to `reverse engineer' the cutoff scheme from 
the Wilsonian action.  In fact, I shall show that  
the Wilsonian action is directly obtained from  
the two-particle-irreducible(2PI) functional.  Since the 2PI functional
has a particular perturbative expansion, one can match the Wilsonian
action with this perturbative expansion, and hence restrict the   
cutoff scheme and field reparametrization that led to the Wilsonian 
action.  For example, if presented with an action and the 
information that it is a Wilsonian effective action, it is possible to
test such assertions in  a straightforward manner.  Quite apart from
these perturbative consistency conditions, the free energy 
interpretation of the Wilsonian action implies non-perturbative convexity
properties, analogous to the usual convexity of the 1PI effective 
action.  In essence, the 2PI formalism enables one 
to phrase cutoff dependence directly in terms of the properties of the
non-perturbative connected two-point function.   

The organization of this paper is as follows:  I first review some
aspects of the 2PI functional, following Cornwall, Jackiw and 
Tomboulis\cite{corn}.  
I then show that the Wilsonian
effective action is the free energy in an ensemble with prescribed 
field expectation value  and {\it prescribed} connected {two-point correlation 
function}.  Since such a free energy is just a particular case of 
the 2PI functional, an infinite number of perturbative consistency 
checks follow immediately from the properties of the 2PI functional.
I  show 
in some  examples how the cutoff is restricted  in this interpretation.
Some comments on the utility of this interpretation are in a 
concluding paragraph. 

\def\ded#1#2{{{\delta {#1}}\over{\delta{#2}}}}
The 2PI functional, in the form I need, was derived by Cornwall, 
Jackiw and Tomboulis\cite{corn} in an unrelated context; 
it was initially studied in by  Lee and 
Yang\cite{leeyang}, and others\cite{others}.
It has been used in the study of the bound state 
problem in quantum field theory, where it can used  to extract
non-perturbative information via variational principles.  
I give a short account of the definition of the 2PI functional, with
some useful properties, in 
Euclidean space, for convenience.  This is essentially taken from
\cite{corn}, replacing various factors 
of $i.$

\def\DD{{\rm D}}
\def\half{\hbox{$1\over 2$}}
If $S_{0}$ is a microscopic action, define a generating functional
\[ W[J,K] \equiv \ln\int \DD\phi \exp(-S_{0}[\phi]+ J \phi + 
\half\phi K \phi) , \]
where   $J \phi \equiv \int dx J(x)\phi(x),$ and $K(x,y)$ is 
a source for the complete two-point function.  
Define the Legendre transform of $W$ 
\begin{equation}
 \Gamma[\varphi,G] + W[J,K] \equiv J \varphi + \half (\varphi 
K\varphi + GK), 
\label{legend}
\end{equation}
($GK\equiv \int dxdy G(x,y)K(x,y)$) which implies that 
\begin{equation}
\ded\Gamma\varphi = J + K\varphi,\qquad \ded\Gamma G = K. 
\label{varyG}
\end{equation}
Since 
\[ \half \langle\phi\phi\rangle =\ded WK = \half(G+\varphi\varphi) ,\]
it follows that $G$ is the exact connected two-point function, and 
$\varphi=\langle\phi\rangle$ as usual.  

 $\Gamma$ is given, in perturbation 
theory, by a sum of graphs that have $G$ as the propagator in internal 
lines, and $\delta^{n}S_{0}[\varphi]/\delta\varphi^{n}$ vertices 
($n>2$).  
Since $G$ is the exact two-point function, there are no
self-energy corrections in this set of graphs; in other words, these
graphs are all 2PI.  There are just two contributions at one-loop 
order,
$\half\int dxdy G(x,y)
{\delta^{2}S_{0}[\varphi]\over\delta\varphi(x)\delta\varphi(y)},$ 
and $-\half\Tr \ln G$ (apart from a term  which is $G$ and $\varphi$
independent).   The formal derivation is given in \cite{corn}.

What does all this have to do with the Wilsonian effective action?
Recall that Polchinski's derivation of the
Wilsonian renormalization group equation\cite{pol} involved the addition of a
cutoff function to 
the bare action.  The cutoff is introduced by adding a term
of the form $\int \half \phi K \phi$ to the action, where the Fourier 
transform of $K$ is taken to vanish for $p^{2}<\Lambda^{2},$ and
$-K/p^{2}$ tends to infinity rapidly for $p^{2}>\Lambda^{2}.$  
(The function $K$ used here is different from $K$ in Polchinski's
paper: $K_{\rm Polchinski}(p^{2}/\Lambda^{2})\equiv (1-K/p^{2})^{-1}.$)
Thus  $W[J,K]$ contains the information needed to 
implement the Wilsonian effective action, for appropriate choices of
$K.$    

However, $K$ is an inherently perturbative object---thinking in terms 
of $K$ requires thinking in terms of Feynman diagrams.  We should 
focus instead on physical quantities, such as $G;$ the awkward 
description of $K$ tending to infinity is expressed as the vanishing 
of $G$ for the appropriate set of modes.  Formally, from 
Eq.~\ref{varyG}, and the usual relation 
\begin{equation}
\Gamma[\varphi] + W[J] = \varphi J
\label{usual} 
\end{equation}
we note that   modes for which $K$ and $J$ vanish have, in fact, been
integrated out---the partition function has been exactly evaluated as
an integral over these modes.  On the other hand,
modes for which $G$ vanishes, 
are uncorrelated in 
a non-perturbative sense.  It is trivial to see that the 
perturbative description  of this non-perturbative characterization 
is that there are no internal lines involving the propagation of 
such modes in Feynman diagrams.  

To see that this is precisely what  the Wilsonian 
effective action requires, let us suppose that we have two fields $\phi, \Phi,$
and we want the Wilsonian effective action for $\phi$ obtained by
integrating out $\Phi.$  Then the claim is that 
\[ S_{W}(\phi) = \Gamma[\phi,\Phi_{s},G_{\phi}=0,G_{\Phi,s}]
\]
where 
\begin{equation}
\Phi_{s}:\ded\Gamma{\Phi_{s}} = 0 ,\qquad G_{\Phi,s}:
\ded\Gamma{G_{\Phi,s}}=0 .
\label{obv}
\end{equation}
Clearly, Eq.~\ref{obv} implies that the integral over $\Phi$ has been
carried out exactly, and $G_{\phi}=0$ is precisely the 
non-perturbative analogue of the statement that
there are no internal lines with $\phi$ contractions.  Note that, in 
general, $\Phi_{s}=\Phi_{s}(\phi), G_{\Phi,s} = G_{\Phi,s}(\phi).$

This identification is pretty much content-free if one could use 
hard cutoffs, with no field dependence.  
Unfortunately, for the applications of interest, one  cannot use such 
simple cutoffs.  $G$ {\it cannot} be taken to vanish abruptly, in which case
one must find the appropriate smooth transition from $G=0$ to 
$G:\ded\Gamma{G}=0,$ which preserves desirable properties such as 
holomorphy.  The point is that $G$ is an object with a non-perturbative 
meaning---it is the connected two-point function in the ensemble of 
interest.  The field dependence of cutoffs can be stated in a precise 
manner now, as $\varphi$-dependence of the transition from
$\ded{\Gamma[\varphi,G]}{G} = 0$ to $G=0.$

As a first  example, suppose that we are given a low-energy effective 
action for a real scalar field at scale $\Lambda_{R}$ in four 
dimensions
and told that 
the mass parameter has no quadratic dependence on $\Lambda_{R}.$  For 
a general microscopic action 
\[ S_{0}[\phi] = \int \half \phi(p^{2}+m^{2})\phi + {g\over 
4!}\phi^{4 }   \]
with microscopic cutoff $\Lambda$ held fixed, this implies that 
\[Z_{m} m^{2}(\Lambda_{R}) = m^{2}+ g \int^{\Lambda} d^{4}p G(p^{2})  \]
where $G=0$ for $p^{2} < \Lambda_{R}^{2}.$  For
$m^{2}(\Lambda_{R})$ to have no quadratic dependence on 
$\Lambda_{R},$ we must have 
\[ \tilde G(x) \equiv xG(x) : 
\int_{0}^{(\Lambda/\Lambda_{R})^{2}} dx x\tilde G^{\prime} = 0.\]
(A similar equation appears in an appendix (eq. A.17) in \cite{hl}, but these 
authors did not make the following observations.)
In other words, the cutoff must be such that the correlation between 
modes being integrated out does not monotonically decay---it {\it must}
oscillate, so that $\tilde G^{\prime}$ is not a positive definite 
function.  It is difficult to see how such a cutoff could be 
consistent with unitarity {\it given} the assumed microscopic dynamics, 
which involves no other degrees of freedom.  Thus, in this example, we 
find that the assumed form of the low-energy Wilsonian action suggests
the existence of other degrees of freedom.  This is reminiscent of 
Intriligator's arguments for `integrating in' degrees of 
freedom\cite{int}.

As a second example, consider an $N=1$ supersymmetric model in
four dimensions, with just chiral multiplets.  In this case, the
interaction terms in a renormalizable microscopic superspace 
action are purely cubic 
and all two-loop and higher contributions to the Wilsonian action 
depend only on $G_{\Phi\bar\Phi},G_{\bar\Phi\bar\Phi},$ and $G_{\Phi\Phi}.$  
It is then simple to see that any transition
to $G_{\Phi\Phi}=0$ that does not violate the tree-level $\theta$ dependence 
will imply the standard superpotential non-renormalization 
theorem\cite{book}.

In conclusion, the characterization of the Wilsonian effective action
given in this note should be useful in restricting higher derivative terms, 
which is of some importance in, {\it e.g.}  
D3-brane scattering\cite{us}.  It should also
be  useful for figuring out potentially interesting
field-dependent regulators for non-supersymmetric situations.  As 
shown in the first example, it may be 
important for ruling out certain regulators as unphysical.   The free 
energy interpretation implies {\it non-perturbative} convexity
properties for the Wilsonian action, to which I hope to return 
elsewhere.


I am grateful to  S. Minwalla, M. Ro\v cek and 
R. von Unge for helpful conversations.
This work was supported in part by NSF grant PHY96-00258.


\begin{references}

\bibitem{wilson} K.G. Wilson and J.B. Kogut,   Phys. Rep. 12C 
 75 (1974)

\bibitem{sw} N. Seiberg and E. Witten, Nucl. Phys. B426, 19 (1994); 
B431, 484 (1994)

 \bibitem{seiberg} N.  Seiberg, Phys.  Lett.  B318, 469 (1993); 
Nucl. Phys. B435, 129 (1995); 
K. Intriligator and N. Seiberg, {  Nucl. Phys. Proc. 
Supp.}{ 45BC}, 1 (1996) 

\bibitem{reviews} For reviews, see M. Peskin, hep-th/9702094; 
L. Alvarez-Gaum\'e and
 F. Zamora, hep-th/9709180 
 
\bibitem{dine} M. Dine and Y. Shirman, Phys. Rev. D50, 1362 (1994); 
S. de Alwis, hep-th/9508053; F. Bastianelli, {  Phys.
Lett.} {B369},  249 (1996)

\bibitem{shif} M. Shifman and A. Vainshtein, Nucl. Phys. B359, 571 
(1991)

\bibitem{corn} J.M. Cornwall, R. Jackiw and E. Tomboulis, 
Phys. Rev. D10, 2428 (1974)

\bibitem{leeyang} T.D. Lee and C.N. Yang, Phys. Rev. 117, 22 (1960)

\bibitem{others} J.M. Luttinger and J.C. Ward, Phys. Rev. 118, 1417 
(1960); P. Martin and C. de Dominicis, J. Math. Phys. 5, 14 (1964);
H.D. Dahmen and G. Jona-Lasinio, Nuovo Cim. 52A, 807 (1967); A.N. 
Vasil'ev
and A.K. Kazanskii, Theor. Math. Phys. 12, 875 (1972)
 
\bibitem{pol} J. Polchinski, {  Nucl. Phys.} {  B231},  269 (1984)

\bibitem{hl} J. Hughes and J. Liu, {  Nucl. Phys.} { B307},
183 (1988)
 
\bibitem{int} K. Intriligator, Phys. Lett. B336, 409 (1994)

\bibitem{book} S.J.~Gates, M.T.~Grisaru, M.~Ro\v cek and W.~Siegel,
{\it Superspace}, (Benjamin-Cummings, Reading MA, 1983)

\bibitem{us} V. Periwal and R. von Unge, hep-th/9801121

\end{references}
\end{document}